\documentclass{elsart4-1}

\usepackage{epsfig}
\usepackage{amssymb}
\usepackage[english,francais]{babel}

\newcommand{\be}{\begin{equation}}
\newcommand{\ee}{\end{equation}}

\def \pmbtext#1{\leavevmode
     \setbox0\hbox{#1}
     \kern-0,2pt \copy0 \kern-\wd0
     \kern0,4pt \copy0 \kern-\wd0
     \kern-0,2pt \raise0,3pt \box0 }

\def\bb0{{\bf B_0}}
\def\rr{{\bf r}}

\def\lam{{\bf e_z}}

\def\eg{{\it e.g.}\ }

\def\bB{{\bf b}}
\def\bv{{\bf v}}
\def\zp{{\bf z^\pm}}
\def\zm{{\bf z^\mp}}

\begin{document}
\centerline{Physics}
\begin{frontmatter}
\selectlanguage{english}
\title{Third-order Els\"asser moments in axisymmetric MHD turbulence}
\selectlanguage{english}
\author[authorlabel1,authorlabel2]{S\'ebastien Galtier}
\ead{sebastien.galtier@ias.u-psud.fr}
\address[authorlabel1]{Univ Paris-Sud, Institut d'Astrophysique Spatiale, UMR 8617, b\^at. 121, F-91405 Orsay, France}
\address[authorlabel2]{Institut universitaire de France}
\medskip
\begin{center}
{\small Received \today; accepted after revision +++++}
\end{center}

\begin{abstract}
Incompressible MHD turbulence is investigated under the presence of a uniform magnetic field $\bb0$. 
Such a situation is described in the correlation space by a divergence relation which expresses the statistical 
conservation of the Els\"asser energy flux through the inertial range. The ansatz is made that the development 
of anisotropy, observed when $B_0$ is strong enough, implies a foliation of space correlation. A direct 
consequence is the possibility to derive a vectorial law for third-order Els\"asser moments which is 
parametrized by the intensity of anisotropy. We use the so-called critical balance assumption to fix this 
parameter and find a unique expression. {\it To cite this article: S. Galtier, C. R. Physique 11 (2010).}

\vskip 0.5\baselineskip

\selectlanguage{francais}
\noindent{\bf R\'esum\'e}
\vskip 0.5\baselineskip
\noindent
{\bf Moments d'Els\"asser du troisi\`eme ordre en turbulence MHD axisym\'etrique.}
La turbulence MHD incompressible est \'etudi\'ee en pr\'esence d'un champ magn\'etique uniforme $\bb0$. 
Une telle situation est d\'ecrite dans l'espace des corr\'elations par une relation de divergence qui exprime 
la conservation statistique du flux d'\'energie d'Els\"asser \`a travers la zone inertielle. Nous faisons l'ansatz 
que l'anisotropie, observ\'ee quand $B_0$ est suffisamment fort, implique un feuilletage de l'espace des 
corr\'elations. Une cons\'equence directe est la possibilit\'e d'obtenir une nouvelle loi vectorielle pour les 
moments d'Els\"asser d'ordre trois qui est param\'etris\'ee par l'intensit\'e de l'anisotropie. Nous utilisons 
l'hypoth\`ese d'\'equilibre critique pour fixer ce param\`etre et trouver une expression unique. 
{\it Pour citer cet article~: S. Galtier, C. R. Physique 11 (2010).}

 \vskip 0.5\baselineskip
\noindent{\small{\it Mots-cl\'es~:} MHD; Turbulence; Vent solaire}}
\end{abstract}
\end{frontmatter}

\selectlanguage{english}

\section{Introduction}
\label{intro}

Despite its large number of applications such as climate, atmospherical flows or space plasmas, 
turbulence is still today one of the least understood phenomena in classical physics; for that reason 
any exact results appear extremely important \cite{frisch}. The Kolmogorov's four-fifths (K41) law 
\cite{K41} is often considered as the most important result in three-dimensional (3D) homogeneous 
isotropic turbulence: it is an exact and nontrivial relation derived from Navier-Stokes equations which 
implies the third-order longitudinal structure function. When isotropy is {\it not} assumed the 
primitive form of the K41 law is the divergence equation \cite{monin} 
\be
-{1 \over 4} \nabla_\rr \cdot {\bf F^{HD}(\rr)} = \varepsilon \, , 
\label{zz1}
\ee
where $\varepsilon$ is the mean energy dissipation rate per unit mass, $\rr$ is the separation vector, 
${\bf F^{HD}(\rr)}= \langle \delta \bv \delta \bv^2 \rangle$ is associated to the energy flux vector and 
$\delta \bv = \bv ({\bf x} + \rr) - \bv ({\bf x})$. Then, the K41 law may be seen as a non trivial 
consequence of equation (\ref{zz1}) when isotropy is assumed; it is written as \cite{K41}
\be
- {4 \over 5} \varepsilon r = \langle \delta v_L^3 \rangle \, , 
\label{zz3}
\ee
where $L$ means the longitudinal direction along $\rr$. 
Few extensions of such a result to other fluids have been made; it concerns \eg scalar 
passively advected such as the temperature or a pollutant in the atmosphere \cite{yaglom} or 
space magnetized plasmas described in the framework of magnetohydrodynamics (MHD) 
\cite{PP98}, electron \cite{Galtier08a} and Hall \cite{Galtier08b} MHD. 

In this paper we investigate 3D homogeneous incompressible MHD turbulence for which the 
following divergence relation holds \cite{PP98}
\be
-{1 \over 4} \nabla_\rr \cdot {\bf F^\pm(\rr)} = \varepsilon^\pm  \, , 
\label{zz5}
\ee
where ${\bf F^\pm(\rr)} = \langle \delta \zm (\delta \zp)^2 \rangle$, $\zp=\bv \pm \bB$ are the Els\"asser 
fields and $\varepsilon^\pm$ are the mean Els\"asser energy dissipation rates per unit mass. 
When isotropy is assumed we obtain the exact law for 3D MHD \cite{PP98}
\be
- {4 \over 3} \varepsilon^\pm r = \langle \delta z^\mp_L (\delta \zp)^2 \rangle \, ,
\label{zz6}
\ee
which may reduce to expression (\ref{zz3}) when the magnetic field is taken equal to zero. 
It is straightforward to demonstrate the compatibility between relations (\ref{zz5}) and (\ref{zz6}) 
by performing an integration of the former over a full sphere (ball). The same remark holds for the 
compatibility between expression (\ref{zz1}) and the K41 law. 

To date the universal isotropic scaling relations discussed above have never been generalized to 3D 
homogeneous -- non isotropic -- turbulence (see however \cite{Galtier09a,Galtier09c} for the latest progress). 
It is basically the goal of this paper to show that an exact relation may be derived in terms of Els\"asser 
fields for axisymmetric MHD turbulence. This derivation is based on the ansatz that the space correlation is 
foliated when the field fluctuations are dominated by a uniform magnetic field. 
Note that a first analysis was made for such a problem in \cite{Galtier09a}. The main goal was the development 
of a tensorial analysis only for vectors $\bv$ and $\bB$ since the Els\"asser fields, a mixture of a vector and 
a pseudo-vector, renders the study much more difficult. Then, the idea of foliation of space correlation was 
eventually introduced to derive a law for third-order correlations in $\bv$ and $\bB$. In the present paper we 
show that the extension of the latter idea to Els\"asser variables is possible -- independently of their tensorial 
nature since we do not perform a tensorial analysis -- and we derive the corresponding exact law.

\section{Impact of a mean magnetic field}
\label{impact}

The influence of a large-scale magnetic field $\bb0$ on the nonlinear MHD dynamics has been 
widely discussed during the last fifteen years. The first heuristic picture of MHD turbulence proposed by 
Iroshnikov-Kraichnan \cite{iro,kraichnan} has been criticized and, nowadays, we know that under 
the presence of $\bb0$ we find turbulent fluctuations with larger fluctuating components in the direction 
transverse to $\bb0$ than along it, as well as different type of correlations along $\bb0$ and transverse 
to it \cite{Galtier00,cho1,Boldyrev,Alexakis07a,Alexakis07b,Bigot1,Bigot2}. 
In other words, the nonlinear transfer occurs differently according to the direction considered with a 
weaker non linear transfer along $\bb0$ than transverse to it, with possibly 
different power law energy spectra. An important concept introduced in the last years is 
the possible existence of a critical balance between the nonlinear eddy-turnover time and the Alfv\'en time 
\cite{gosri}. The former time may be associated to the distortion of wave packets whereas the latter may 
be seen as the duration of interaction between two counter-propagating Alfv\'en wave packets. A direct 
consequence of the critical balance is the existence of a relationship (in the inertial range) between 
length-scales along ($\parallel$) and transverse ($\perp$) to the mean magnetic field direction (see also 
\cite{Galtier05}). This relation, generally written in Fourier space, is
\be
k_\parallel \sim k_\perp^{2/3} \, .
\label{cb}
\ee
In practice, numerical evidences of relation (\ref{cb}) may be found by looking at the parallel and 
perpendicular (to the mean magnetic field direction) 
intercepts of the surfaces of constant energy, either in physical space with 
second-order correlation functions \cite{cho1,Maron} or in Fourier space with spectra \cite{Bigot2}. 
Note that one generally takes a local definition for $k_\parallel$ by using the local mean magnetic field 
but it has been shown that a global definition (with the parallel direction along $\bb0$) works quite well if 
$B_0$ is strong enough \cite{Bigot2}. 
Despite the limitation of direct numerical simulations a scaling relation between parallel and perpendicular 
length scales seems to emerge whose power law relation is compatible with the critical balance relation 
(\ref{cb}). Therefore, the idea of a general relationship between length scales during the nonlinear transfer 
(of energy) from large to small scales may be seen as a natural constrain for theoretical models. 
Basically, we translate this constrain as an ansatz for axisymmetric MHD turbulence which allows us to 
derive from equation (\ref{zz5}) the equivalent of the four-fifths law. 

At this level of discussion, it is interesting to remark that the assumption of isotropy made to derive 
the exact law (\ref{zz6}) is questionable in the sense that we never observe exactly isotropy. For example in 
\cite{milano,ng2} it was shown numerically that despite the absence of a uniform magnetic field ($B_0=0$) 
deviations from isotropy are observed locally with the possibility to get a scaling relation between length-scales 
along and transverse to the local magnetic field. This local anisotropy is expected to be stronger at larger (magnetic) 
Reynolds numbers for which the exact law (\ref{zz6}) is derived. Therefore, this exact law (\ref{zz6}) should be 
seen as a first order description of MHD turbulence when $B_0=0$. More precisely in the derivation of this law 
one should consider the decomposition
\be
{\bf F^\pm(\rr)} = {\bf F_{iso}^\pm}(\rr) + \delta {\bf F_{ani}^\pm}(\rr) \, , 
\ee
where the first term in the RHS is the isotropic contribution to the vector third-order moment whereas the second 
term measures the deviation from isotropy. When the second term is of second order in importance then 
$\delta {\bf F_{ani}^\pm} \ll {\bf F_{iso}^\pm}$ and the integration of relation (\ref{zz5}) over a full sphere -- with 
the application of the divergence theorem -- gives the universal law (\ref{zz6}).

The derivation of a universal law from equation (\ref{zz5}) in the general case of non isotropic turbulence 
is far from obvious. For example, one needs to find a volume ${ V}$ such that at its surface 
${ S}$ the normal component $F_n$ of {\bf F} is conserved.
Then, one can perform an integration of equation (\ref{zz5}) over this volume, apply the divergence theorem 
and obtain a simple expression independent of any parameter. In practice, that means one starts with 
\be
-{1 \over 4} \int\int\int_{ V} \nabla_\rr \cdot {\bf F^\pm(\rr)} d{ V} =
\varepsilon^\pm  \int\int\int_{ V} d{ V} \, , 
\ee
which gives by the divergence theorem and after integration over the volume 
\be
-{1 \over 4} \int\int_{ S} {\bf F^\pm(\rr)} \cdot d{\bf { S}} = \varepsilon^\pm \, { V} \, , 
\ee
and after projection on the surface vector $d{ S}$ 
\be
-{1 \over 4} \int\int_{ S} F_n^\pm(\rr) d{ S} = \varepsilon^\pm \, { V} \, .
\ee
If one assumes that $F_n^\pm(\rr)$ is constant on ${ S}$ then one obtains
\be
-{1 \over 4} F_n^\pm(\rr) \int\int_{ S} d{ S} = -{1 \over 4} F_n^\pm(\rr) { S} 
= \varepsilon^\pm \, { V} \, ,
\ee
which leads to the exact law 
\be
F_n^\pm(\rr) = - 4 \varepsilon^\pm \, {{ V} \over { S}} \, . 
\ee

The form (and even the existence) of such a volume ${ V}$ is still an open question. 
However, it is important to note that there exists an infinity of mathematical solutions of equation (\ref{zz5}) 
but they depend on parameters which render the solutions non universal. For example we may have 
\cite{podesta07} 
\be
{\bf F}^{\pm}(\rr) = -4 \varepsilon (A^{\pm} \rho \, {\bf e_\rho} + (1-2A^{\pm}) z \, {\bf e_z}) \, , 
\label{exF}
\ee
where $\rho$ and $z$ are the cylindrical coordinates, and ${\bf e_\rho}$ and ${\bf e_z}$ are the 
corresponding unit vectors (with $\lam \equiv \bb0 /B_0$). Note that the choice $A^\pm=1/2$ gives 
the universal law for two-dimensional isotropic MHD turbulence, whereas $A^\pm=1/3$ leads to a radial 
vector and corresponds to the three-dimensional isotropic law \cite{PP98}. Then, we may expect that 
relation (\ref{exF}) describes correctly anisotropic MHD turbulence when $A^\pm \in [1/3;1/2]$ with stronger 
anisotropy when $A^\pm$ is closer to $1/2$. However, relation (\ref{exF}) does not satisfy the critical 
balance relation (\ref{cb}) for any values of $A^\pm$: indeed, for isotropic turbulence the energy flux 
vector is radial which may express the fact that energy cascades radially, whereas when a mean magnetic 
field is present it is not the case anymore and iso-contours of spectral energy are elongated in the 
perpendicular direction according to the power law (\ref{cb}) with an elongation more pronounced at small 
length scales (which means, in the correlation space, an elongation along the mean magnetic field 
direction). According to relation (\ref{exF}), we see that for a given distance $r$ the energy flux ratio 
between a point along ${\bf e_z}$ and another point along ${\bf e_\rho}$ is equal to the following constant
\be
{F^{\pm}(r{\bf e_z}) \over F^{\pm}(r{\bf e_\rho})} = {1-2A^{\pm} \over A^{\pm}} \, .
\label{cst}
\ee
This constant can be very small (when $A^\pm$ is close to $1/2$) but its precise value does not change the 
nature of the relation between these two fluxes which is linear. Therefore, it can only lead to a linear law 
dependence between the parallel and perpendicular intercepts of the surfaces of constant energy (the form 
of these surfaces being directly related to the intensity and direction of the energy flux). 
Note that if one considers a slightly 
different situation with points close to the ${\bf e_\rho}$ and ${\bf e_z}$ directions with energy fluxes 
$F^{\pm}(r{\bf e_\rho}+\epsilon{\bf e_z})$ and $F^{\pm}(\epsilon {\bf e_\rho} + r{\bf e_z})$ respectively 
(where $\epsilon$ is a small parameter), the conclusion does not change drastically as long as 
$r \gg \epsilon$; when $r$ becomes of the order of $\epsilon$ then both energy flux vectors deviate 
significantly from the ${\bf e_\rho}$ and ${\bf e_z}$ directions which does not help for increasing anisotropy
at small length scales which needs to have energy flux vectors preferentially along ${\bf e_\rho}$.
Expression (\ref{exF}) is the simplest solution among an infinity of axisymmetric solutions obtained by 
\cite{podesta07}. The expression that we shall derive here for the energy flux vector is another particular 
solution of this family which satisfies this time the critical balance assumption. 

In order to recover an anisotropic law of the type of (\ref{cb}) -- which is a power law -- it is necessary to 
reinforce the energy flux in the ${\bf e_\rho}$ direction at small length scales. Then, the following statement 
is made that the energy flux vector has an orientation closer to the ${\bf e_\rho}$ direction when the length 
scale decreases. This variation must have a power law dependence (with power law index $n$) in the 
length scale in order to be compatible with relation (\ref{cb}) which is also a power law. The value of $n$ 
compatible with the index $2/3$ in relation (\ref{cb}) may be determined with critical balance arguments 
(see Section \ref{sec4}). We will see that if we incorporate such a requirement in the analysis then we may 
derive a universal law in the sense that it does not depend on any (non physical) parameter. In practice, the 
energy flux vectors will belong to an axisymmetric surface ${ S}_n$ in the three-dimensional space 
correlation (which means that ${\bf F}^{\pm}(\rr)$ is tangent to ${ S}_n$ for any points 
$M^\prime \in { S}_n$; see Section \ref{secfole} and Fig. \ref{Fig1}). The manifold 
${ S}_n$ is defined in such a way that the energy flux vectors tend to be perpendicular to ${\bf e_z}$ 
when the distance separation goes to zero which means that turbulence tends to be bi-dimensional at small 
scales. As we will see in Section \ref{sec51}, the expected constant $-2$ for two-dimensional MHD turbulence 
is indeed recovered from the exact law when the small scale limit is taken.

\section{Foliation of space correlation}
\label{secfole}

\begin{figure}[h]
\centerline{\hbox{\psfig{figure=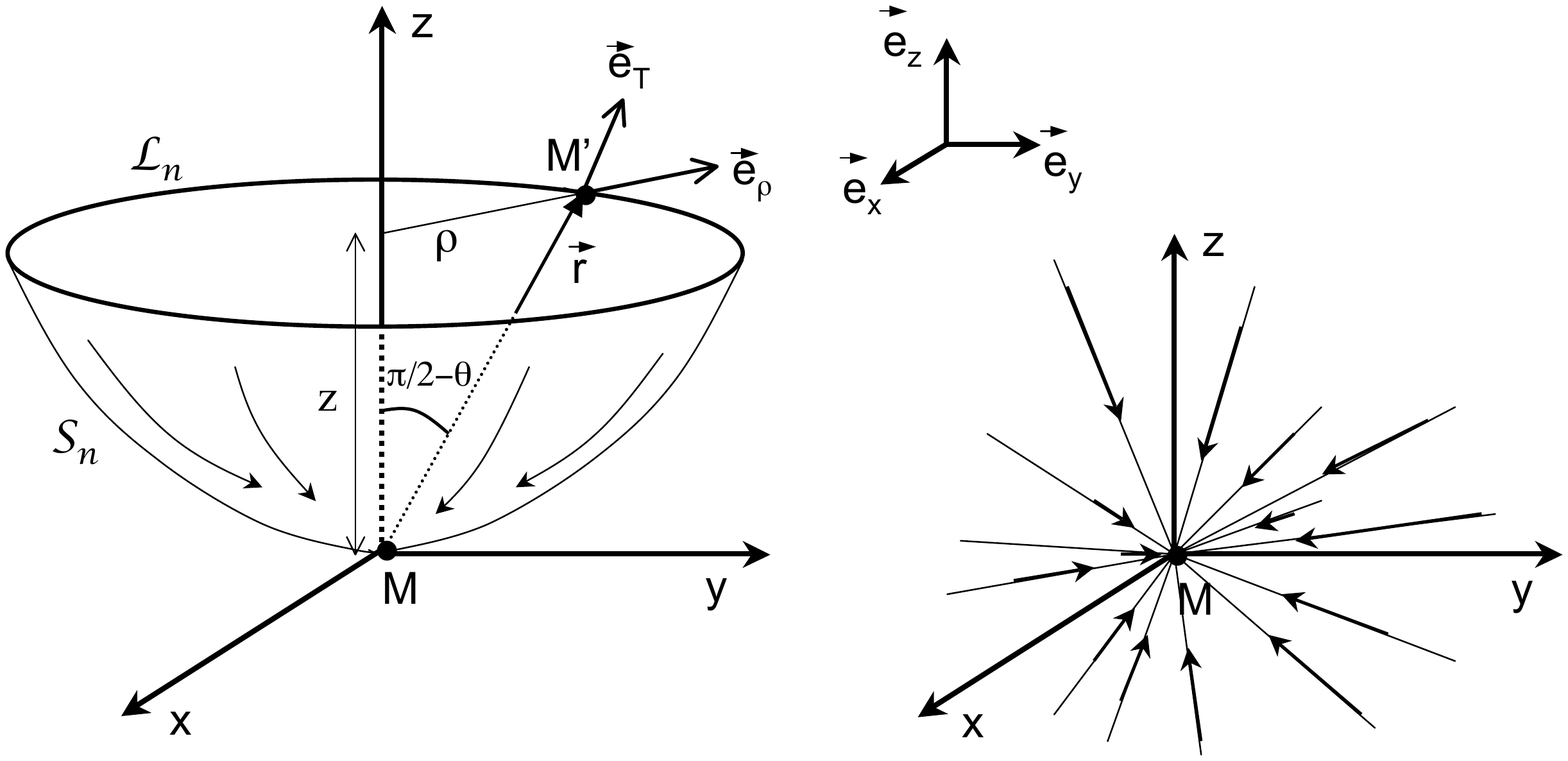,width=14cm,height=8cm}}}
\caption{Left: We perform an integration of relation (\ref{zz5}) over the manifold ${ S}_n$ defined 
in the half upper space by the function $f_n(\rho) = \rho_0 (\rho / \rho_0)^n$ with $n>1$; note the use 
of the polar coordinates with $\rr=(\rho,z)$. ${ S}_n$ is a surface of revolution about the $(Mz)$ axis: 
on this Figure it appears as a "bowl" of axis of symmetry ($Mz$). The vector ${\bf e_T}$ at point 
$M^\prime$ is tangent to the surface ${ S}_n$ and perpendicular to the circle ${ L}_n$ of radius 
$\rho$ which has also ($Mz$) for axis of symmetry. The curved vectors represent schematically the 
orientation of the energy flux ${\bf F^\pm(\rr)}$ which flows towards the point $M$. For comparison the 
three-dimensional isotropic case is also represented (right) for which the energy flux flows radially.}
\label{Fig1}
\end{figure}

From several theoretical and numerical analyses we know that MHD turbulence under the influence of 
$\bb0$ develops anisotropy that increases as the length scale decreases. Additionally, the rms 
fluctuations at a given separation distance $r$ are more intense when ${\bf r}$ is perpendicular to $\bb0$ 
than when ${\bf r}$ is parallel to it. This property can be understood as a consequence of the critical 
balance relation (\ref{cb}) which provides a relationship between the length scales of the fluctuations 
parallel and perpendicular to the mean magnetic field. Following these considerations and those 
exposed at the end of Section \ref{impact}, we make the ansatz that the energy flux vectors belong to 
two-dimensional surfaces ${ S}_n$ in the three-dimensional space correlation (which means that 
${\bf F}^{\pm}(\rr)$ is tangent to ${ S}_n$ for any points $M^\prime \in { S}_n$; see Fig. \ref{Fig1}). 
Since the problem is axisymmetric, the manifolds ${ S}_n$ must be of revolution about the $(Mz)$ 
axis (with $\lam \equiv \bb0 /B_0$; see Fig. \ref{Fig1}). 
It is defined in such a way that the direction of ${\bf F}^{\pm}(\rr)$ tends to become perpendicular to $\lam$ 
when the distance separation $r$ goes to zero. This variation of direction for ${\bf F}^{\pm}(\rr)$ should 
have a power law dependence in the length scale. Then, the axisymmetric manifold ${ S}_n$ is defined 
by the following function
\be
z = f_n(\rho) = \rho_0 \left({\rho \over \rho_0}\right)^n \, . 
\label{fole}
\ee
It is the simplest algebraic function satisfying the conditions $f_n(\rho) \to 0$ when $\rho \to 0$ 
with a simple power law dependence between $\rho$ and $z$. 
Without loss of generality we may already note that $n$ must be greater than one to satisfy the 
anisotropic property (the energy flux vector getting perpendicular to $\bb0$ at small separation 
distance $r$).
Finally, note that $\rho_0$ is the value of $\rho$ for which the angle between $\rr$ and $\lam$ is 
$\pi/4$; therefore $\rho/\rho_0$ may be seen as a way to delimit the correlation space into two domains 
where the direction of the separation vector $\rr$ is closer to the transverse plane ($xMy$) or to 
the parallel direction ${\bf e_z}$ (see Fig. \ref{Fig1}). 
It is important to emphasize that the critical balance measured in MHD turbulence 
(with $B_0>0$) is a situation towards which the nonlinear dynamics converges: it is the main state of 
the dynamics. In other words, deviations from this state may be found but are of second order in magnitude. 
In the same way, the assumption of a foliation of the space correlation (with relation (\ref{fole})) means that 
one should write
\be
{\bf F^\pm(\rr)} = {\bf F_{fol}^\pm}(\rr) + \delta {\bf F_{nonfol}^\pm}(\rr) \, , 
\ee
where the first term in the RHS is the vector third-order moment which belongs to the foliated space 
correlation (see the schematic vectors in Fig. \ref{Fig1}) 
whereas the second term corresponds to other vector contributions which are assumed (ansatz) 
of second order in importance, namely $\delta {\bf F_{nonfol}^\pm} \ll {\bf F_{fol}^\pm}$.

Equation (\ref{zz5}) is integrated over the manifold ${ S}_n$ of axis of symmetry $(Mz)$. 
An illustration is given in Fig. \ref{Fig1} where ${ S}_n$ appears as a "bowl". It gives 
\begin{eqnarray}
- 4 \varepsilon^\pm \int\int_{{ S}_n} d {{ S}_n} &=& \int\int_{{ S}_n} \nabla_\rr \cdot  
{\bf F^\pm(\rr)} \, d{{ S}_n} \, . 
\end{eqnarray}
By the Green's flux theorem (see Appendix) and after integration over the surface, we obtain
\begin{eqnarray}
- 4 \varepsilon^\pm {{ S}_n} &=& \oint_{circle} {\bf F^\pm(\rr)} \cdot d{{ L}_n} \, , 
\end{eqnarray}
where the line integral is performed along a circle ${ L}_n$ of radius $\rho$ and of axis of symmetry 
$(Mz)$. On the example given in Fig. \ref{Fig1}, it corresponds to the upper boundary of the "bowl".
Note that $d{{ L}_n}$ is an elementary vector which is normal to the circle ${ L}_n$ and tangent 
to the surface ${{ S}_n}$ (see Appendix). Then, one gets after projection
\begin{eqnarray}
- 4 \varepsilon^\pm {{ S}_n} &=& \oint_{circle} F_T^\pm(\rr) d{{ L}_n} \, ,
\end{eqnarray}
where $T$ means the tangent direction at point $M^\prime$ (see Fig. \ref{Fig1}). 
The problem being axisymmetric, $F_T(r)$ is unchanged along the circle ${{ L}_n}$ 
of axis of symmetry $(Mz)$; then we have 
\begin{eqnarray}
- 4 \varepsilon^\pm {{ S}_n} &=& F_T^\pm(\rr) \oint_{circle} d{{ L}_n} = F_T^\pm(\rr) \, 2 \pi \rho \, ,
\end{eqnarray}
and thus
\be
- {4 \varepsilon^\pm {{ S}_n} \over 2 \pi \rho} = F_T^\pm(\rr) \, . 
\label{stnew3}
\ee
If we introduce the unit vector ${\bf e_T}$ along the T--direction we obtain the vectorial relation 
\be
- {2 \varepsilon^\pm {{ S}_n} \over \pi \rho} \, {\bf e_T} = {\bf F_T^\pm}(\rr) \, ,
\label{stnew4}
\ee
with
\begin{eqnarray}
{\bf e_T} &=& {{\bf e_\rho} + f_n^\prime (\rho) \lam \over \sqrt{1 + {f_n^\prime(\rho)}^2}}
= {{\bf e_\rho} + n (\rho / \rho_0)^{n-1} \lam \over \sqrt{1 + n^2 (\rho / \rho_0)^{2(n-1)}}} 
= {{\bf e_\rho} + n \tan \theta \lam \over \sqrt{1 + n^2 \tan^2 \theta}} \, ,
\end{eqnarray}
where $\theta$ is the angle between $\rr$ and the ($xMy$) plane (see Fig. \ref{Fig1}). 
Note that for the foliated space correlation defined with relation (\ref{fole}) the general form of the 
divergence operator is 
\be 
\nabla \cdot {\bf F} \equiv {1 \over \rho} {\partial (\rho F_T) \over \partial T} + {1 \over \rho} 
{\partial F_\phi \over \partial \phi} \, , 
\ee
where $\phi$ is the angle defined in cylindrical coordinates (note that by symmetry $F_\phi=0$) and $dT$ is 
the unit length along the tangent direction (see Fig. \ref{Fig1}). The surface ${{ S}_n}$ for a given $\rho$ 
is defined as
\begin{eqnarray}
{{ S}_n} &=& \int 2 \pi \rho \, dT
= \int_0^\rho 2 \pi \rho \sqrt{1 + {f_n^\prime(\rho)}^2} d\rho 
= \int_0^\rho 2 \pi \rho \sqrt{1 + n^2 \left( {\rho \over \rho_0} \right)^{2(n-1)}} d\rho \nonumber \\
&=& {\pi \rho_0^2 \over n^{2/(n-1)}}  \int_0^{X} \sqrt{1 + X^{n-1}} dX \, , 
\end{eqnarray}
with 
\be
X = n^{2/(n-1)} \left(\rho \over \rho_0\right)^2 = \left( n z \over \rho \right)^{2/(n-1)} 
= \left( n \tan \theta \right)^{2/(n-1)} \, .
\label{relationX}
\ee
The combination of the different expressions gives eventually the following vectorial law for Els\"asser fields
\be
- 2 {I(X) \over X} \varepsilon^\pm \rho \, {\bf e_T} = {\bf F_T^\pm}(\rr) \, ,
\label{aniso_v1}
\ee
where 
\be
I(X) = \int_0^{X} \sqrt{1 + X^{n-1}} dX \, .
\ee

\section{Critical balance condition}
\label{sec4}

The vectorial relation (\ref{aniso_v1}) implies a parameter $n$ that has to be determined. 
We shall fix $n$ by a dimensional analysis based on the critical balance condition \cite{gosri}. 
To investigate this idea we will restrict our analysis to the inviscid, stationary MHD equations since 
basically we want an interpretation of relation (\ref{aniso_v1}) valid in the inertial range; we thus obtain
\be
\zm \cdot \nabla \, \zp = - {\bf \nabla} P_* \pm B_0 \partial_\parallel \zp \, , 
\label{mhdb1} 
\ee
where $P_*$ is the total pressure. 
By first noting that the divergence operator applied to (\ref{mhdb1}) allows us to link the total
pressure to the left hand side term, and second that $z^+ \sim z^-$ for small cross-correlation; 
we then arrive to the nontrivial critical balance 
\be
z^\pm_r \nabla_r \sim B_0 \partial_\parallel \, ,
\ee
which may also be written as
\be
{z^\pm_r \over B_0} \sim {\partial_\parallel \over \nabla_r} \sim {k_\parallel \over k_r} = \sin \theta \, ,
\label{anglez}
\ee
where $\theta$ is also the angle between the separation vector $\rr$ and the ($xMy$) plane 
(see Fig. \ref{Fig1}). As we see, relation (\ref{anglez}) offers a direct evaluation of the $\rr$--direction: 
therefore, although the external magnetic field does not enter explicitly in the vectorial relation 
(\ref{aniso_v1}), it constrains -- as expected -- the direction along which the scaling law applies. 
If we now come back to relation (\ref{aniso_v1}), we may write (at first order for small length scales) 
the dimensional relation which is independent of $n$
\be
z^\pm_r \sim (\varepsilon^\pm \rho)^{1/3} \, , 
\label{zozo2}
\ee
and obtain
\be
\sin \theta \sim {(\varepsilon^\pm \rho)^{1/3}  \over B_0} \, .
\label{anglezz}
\ee
In other words, this result means that the scaling relation depends on the strength of the external 
magnetic field with an orientation close to the ($xMy$) plane for strong 
$B_0$, but also on the scales itself with a direction getting closer to the ($xMy$) plane 
at small scales (small $r$). This dimensional analysis will be used below to derive the unique expression 
of the vectorial law for anisotropic MHD turbulence since relation (\ref{anglezz}) gives the following 
dimensional small-scale constraint 
\be
\sin \theta \sim {(\varepsilon^\pm \rho)^{1/3}  \over B_0} \sim \left({\rho \over \rho_0}\right)^{n-1} \, , 
\ee
which leads to $n=4/3$. 
Note that for other types of fluids the value of $n$ may be different \cite{Galtier09c}. 

\section{Exact vectorial law}

Following the critical balance idea we shall rewrite expression (\ref{aniso_v1}) for $n=4/3$ 
which gives
\be
- g(\theta) \varepsilon^\pm r \, {\bf e_T} = {\bf F_T^\pm}(\rr) \, ,
\label{aniso_v2}
\ee
with $g(\theta) \equiv 2 \cos \theta I(X)/X$,
\be
X = \left( {4 \over 3} \tan \theta \right)^6 , \, \, 
{\bf e_T} = {{\bf e_\rho} + (4/3) \tan \theta \lam \over \sqrt{1 + (4/3)^2 \tan^2 \theta}} \, ,
\ee
and
\begin{eqnarray}
I(X) &=& \int_0^X \sqrt{1 + X^{1/3}} \, dX \\
&=& -{16 \over 35} + {6 \over 7} \left( 1 + X^{1/3} \right)^{3/2} X^{2/3} 
- {24 \over 35} \left( 1 + X^{1/3} \right)^{3/2} X^{1/3} + {16 \over 35} \left( 1 + X^{1/3} \right)^{3/2}  \, . \nonumber
\end{eqnarray}
It is the final form of the exact law. We see that the vectorial law has a form close to the 
isotropic case (\ref{zz6}) with a scaling linear in $r$. However, we observe a $\theta$-angle dependence 
which reduces the degree of universality of the law. From an observational point of view this prediction 
turns out to be interesting since in the solar wind the measurements are naturally made at a given 
angle. Numerical estimate of the function $g(\theta)$ gives a slight variation from $2$ to $16/7$ for respectively 
$\theta=0$ to $\pi/2$. It is important to remark that this law is valid for {\it any} $r$ and $\theta$ which means that 
we may describe the entire correlation space. 
Note that the law derived here implies only the mean Els\"asser energy dissipation rates per unit mass 
$\varepsilon^\pm$ which makes a difference with other types of universal results like in wave turbulence where 
the spectra may be expressed in terms of directional energy fluxes (like $P^\pm_\perp$ or 
$P^\pm_\parallel$) \cite{Galtier06,Galtier09r}.


\section{The two-dimensional limit}
\label{sec51}

It is interesting to analyze the small $\theta$ limit for which the energy flux vector is mainly transverse. 
For this limit, we obtain after a Taylor expansion
\be
I(X) \simeq X + {3 \over 8} X^{4/3} \, , 
\ee
and then after substitution
\be
- 2 \left(1 + {2 \over 3} \tan^2 \theta \right) \varepsilon^\pm \rho \, {\bf e_T} 
\simeq {\bf F_T^\pm}(\rr) \, .
\label{aniso_vt}
\ee
This relation tends asymptotically to the scaling prediction for 2D MHD turbulence which may be obtained 
directly after integration (and application of the Green's flux theorem) of expression (\ref{zz5}) over a disk 
with only transverse fluctuations. This result shows in particular how close we are from a two-dimensional 
turbulence.

\section{Discussion and conclusion}

The interplanetary medium is probably the best example of application of the new exact law in terms of 
Els\"asser fields. Indeed, it is a medium permeated by the solar wind, a highly turbulent and anisotropic 
flow which carries the solar magnetic field \cite{klein,Bieb}. Several recent works have been 
devoted to the analysis 
of low frequencies solar wind turbulence in terms of structure functions by using the exact isotropic law 
\cite{sorriso}. A direct evidence for the presence of an inertial energy cascade in the solar 
wind is claimed but the comparison between data and theory is moderately convincing because of 
the narrowness of the inertial range measured. Some recent improvements have been obtained by 
using a model of the isotropic law where compressible effects are included \cite{carbone}. Even if the 
result seems to be better the hypothesis of isotropy is a serious default. Other applications of the 
MHD laws (exact or modeled) are also found in order for example to evaluate the local solar wind heating 
\cite{marino} along or transverse to the mean magnetic field.

Direct numerical simulations are very important to check for example the applicability of the universal 
laws discussed in the present paper since there are exact as long as the hypotheses are satisfied. 
For example, in the isotropic case it is interesting to note that the constant has never been checked 
-- only the power law. Therefore, we are not yet at the same degree of achievement reached for the 
four-fifth's law for which the constant has been recovered experimentally \cite{antonia}. Then, for the 
exact vectorial law derived in this paper it is fundamental to check not only the power law dependence 
(actually, a first analysis at moderate numerical resolution of $256^3$ shows a relatively good agreement 
with the scaling prediction) but also -- and more importantly -- the coefficient $g(\theta)$ which is around 
$2$. Only massive numerical simulations like in \cite{mininni} will allow to take up this challenge.

The interplanetary medium is an excellent laboratory to test new ideas in turbulence. In that respect, 
it would be interesting to extend the present work to other invariants like the cross-correlation. Recent 
works have been devoted to this problem where the idea of a dynamic alignment between the velocity 
and the magnetic field fluctuations has emerged \cite{Boldyrev} but the confrontation with solar wind 
data is still not totally convincing \cite{podesta08}. 
Since most of astrophysical space plasmas evolve in a medium where a magnetic field is present on 
the largest scale of the system the present law has potentially a lot of other applications.

\section*{Acknowledgment}
I acknowledge Institut universitaire de France for financial support.

\appendix

\section{Green's flux theorem}

\begin{figure}[h]
\centerline{\hbox{\psfig{figure=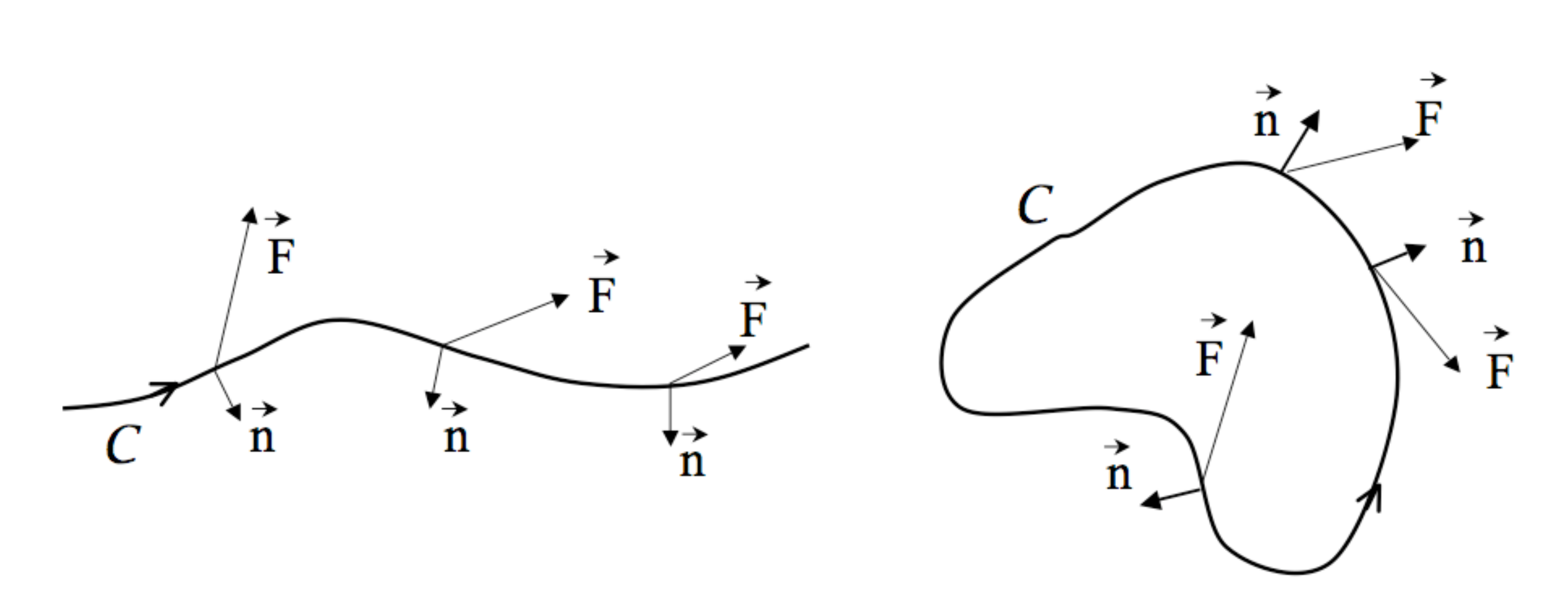,width=11cm,height=3.4cm}}}
\caption{Left: Oriented plane curve ${ C}$ across which the flux of ${\bf F}$ is  computed. 
The normal direction is oriented $90$ degrees clockwise from the tangent direction. 
Right: Oriented plane curve ${ C}$ that encloses a region $S$.}
\label{Fig3}
\end{figure}

The appendix is devoted to the Green's flux theorem which may be seen as the two-dimensional 
version of the well-known divergence theorem. It is also called the Normal form of Green's theorem. 
Let us consider an oriented plane curve ${ C}$ and a plane vector field ${\bf F}$ defined along 
${ C}$. Then the flux of ${\bf F}$ across ${ C}$ is the line integral 
\be
\int_{ C} {\bf F} \cdot {\bf n} \, d \ell \, , 
\ee
where ${\bf n}$ is the unit vector normal to the curve ${ C}$ pointing $90$ degrees clockwise from the 
tangent direction of ${ C}$ (see Fig. \ref{Fig3}; left) and $d\ell$ is an elementary length of curve ${ C}$. 

If now ${ C}$ is a curve that encloses a region $S$ counterclockwise (see Fig. \ref{Fig3}; right) and 
if ${\bf F}$ is defined in the plane (on ${ C}$ and also in $S$), then we have the relation
\be
\oint_{ C} {\bf F} \cdot {\bf n} \, d \ell = \int \int_S {\bf \nabla} \cdot {\bf F} \, dS \, ,
\label{B2}
\ee
which means that the flux of ${\bf F}$ across a closed integral line is equal to the sum of the divergence of 
${\bf F}$ on the surface $S$. It is the Green's flux theorem. 

A short proof of the Green's flux theorem comes as follows. Let us consider the particular case of a rectangular 
closed curve ABCDA whose orientation defines the x and y directions. On the 
one hand, one has 
\begin{eqnarray}
\oint_{ C} {\bf F} \cdot {\bf n} \, d\ell &=& \oint_{ C} {F_x \brace F_y} \cdot {dy \brace -dx} 
= - \int_A^B F_y(x,y_1) dx + \int_B^C F_x(x_2,y) dy - \int_C^D F_y(x,y_2) dx + \int_D^A F_x(x_1,y) dy \nonumber \\
&=& - \int_{x_1}^{x_2} (F_y(x,y_1) - F_y(x,y_2) ) dx + \int_{y_1}^{y_2} (F_x(x_2,y) - F_x(x_1,y) ) dy \, .
\end{eqnarray}
On the other hand, one has
\begin{eqnarray}
\int \int_S {\bf \nabla} \cdot {\bf F} \, dS &=& \int_{x_1}^{x_2} \int_{y_1}^{y_2} (\partial_x F_x + \partial_y F_y) dxdy \\
&=& \int_{y_1}^{y_2} (F_x(x_2,y)-F_x(x_1,y)) dy + \int_{x_1}^{x_2} (F_y(x,y_2)-F_y(x,y_1)) dx \, ,
\end{eqnarray}
which is equal to the flux.


\end{document}